\documentclass[twocolumn,showpacs,prc]{revtex4}
\usepackage{graphicx}
\usepackage{dcolumn}
\usepackage{bm}
\begin{document}

\title{Entropy and the Cosmic Ray Particle Energy Distribution Power Law Exponent}
\author{A. Widom and J. Swain}
\affiliation{Physics Department, Northeastern University, Boston MA USA}
\author{Y.N. Srivastava}
\affiliation{Physics Department, University of Perugia, Perugia IT}

\begin{abstract}
We consider the hypothesis that cosmic rays are emitted from the surfaces of 
neutron stars by a process of evaporation from an internal nuclear liquid to 
a dilute external gas which constitutes the ``vacuum''. On this basis, we find 
an inverse power in the energy distribution with a power law exponent of 
2.701178 in excellent agreement with the experimental value of 2.7. The 
heat of nuclear matter evaporation via the entropy allows for the computation 
of the exponent. The evaporation model employed is based on the entropy 
considerations of Landau and Fermi that have been applied to the liquid drop 
model of evaporation in a heavy nucleus excited by a collision. This model
provides a new means of obtaining power law distributions for cosmic ray
energy distributions and, remarkably, an actual value for the exponent which
is in agreement with experiment and explains the otherwise puzzling smoothness
of the cosmic ray energy distribution over a wide range of energies without
discontinuities due to contributions from different sources required by current
models.
\end{abstract}

\pacs{13.85.Tp, 13.85.Dz, 13.85.Lg}

\maketitle

\section{Introduction \label{intro}}

The classic treatises on sources of cosmic ray energy
distributions\cite{Gaisser: 1990, Stanev: 2004, Auger: 2013}
discuss a power law probability rule whose theoretical
basis is presently not entirely clear\cite{Gaisser:2002}.
A variety of mechanisms have been proposed, assuming
either top-down, where initially very high energy particles
come from decays of heavy remnants from the early universe
and bottom-up ones which involve cosmic accelerators
of various kinds, including ``one-shot'' acceleration, perhaps
around neutron stars or black holes where large electric
fields might be found with relatively small orthogonal
magnetic fields which would otherwise produce energy
losses due to synchrotron radiation, and diffusive
(Fermi) acceleration via collisions
of charged particles with moving magnetic fields due
to astrophysical shock waves.

Confining attention to the region in which the observed
differential flux with respect to energy $dN/dE$ is proportional
to $E^{-\alpha}$ with $\alpha$ near 2.7, there are at least two notable problems with traditional explanations.
The first is obtaining the exponent 2.7. While Fermi acceleration can at least give rise
to a power law spectrum, it is difficult
to argue for the correct exponent. In first order Fermi acceleration
the exponent is 1 plus a source-dependent correction, while 
second order Fermi acceleration gives an (incorrect) 
exponent of 2. A short and very readable account of 
high energy cosmic ray acceleration mechanisms can be
found in \cite{Bustamante}.

Secondly, since multiple sources would contribute
to what is measured on earth, one has to explain: 1) why the observed
spectrum is as smooth as it is ({\em i.e.} why each source has,
within observed errors, the same exponent),  and 2) why various 
sources should be distributed in such a way and with such
intensities that there would, even with the same slopes, be
no jumps in the spectrum. 

This suggests that a more economical
explanation not dependent on such apparent fine-tuning, and,
if possible, giving the correct exponent, would
be worth considering.

A statistical thermodynamic view of the energy distribution 
was pioneered employing a non-extensive 
entropy\cite{Tsallis:1988,Tsallis:1998,Tsallis:2004,Wilk:2000} 
and later interpreted in terms of temperature 
fluctuations\cite{Beck:2001,Beck:2003,Beck:2004,Beck:2005}. 
For our purposes we apply {\em entropy computations} in a form 
originally due to Landau\cite{Landau:1959} for the Landau-Fermi  
liquid drop model of a heavy nucleus. The energy 
distribution of some of the decay products are thought to be 
evaporating nucleons from the bulk 
liquid drops excited by a heavy nuclear collision\cite{Blatt:1979}. 
Although the entropy for this model is computed from 
non-relativistic quantum mechanics, the cosmic ray version 
of evaporation of course requires an ultra-relativistic limit.

Our purpose is to explain the empirical measured scaling law 
index \begin{math} \alpha  \end{math} which appears in the 
energy distribution law of cosmic ray nuclei via the 
differential flux 
\begin{equation}
\left[\frac{d^4 \bar{N}}{dt dA d\Omega dE }\right]\approx 
\frac{1.8 \ {\rm Nucleons}}{{\rm sec\ cm^2\ sr\ GeV}} 
\left( \frac{1\ {\rm GeV}}{E} \right)^\alpha   
\label{intro1}
\end{equation}
that holds experimentally true\cite{Groom:2000} for the 
energy interval 
\begin{math} 5{\rm \ GeV }<E< 100 {\rm \ TeV } \end{math}.
Our theoretical result is that 
\begin{equation}
\alpha = 3\left[\frac{\zeta(4)}{\zeta(3)}\right] 
\label{intro2}
\end{equation}
wherein
\begin{equation}
\zeta(s)=\sum_{n=1}^\infty \frac{1}{n^s}  
\label{intro3}
\end{equation}
is the Riemann zeta function. Numerically, our theoretical 
value is 
\begin{equation}
\alpha_{\rm th} \approx 2.701178 \ \ \ ({\rm theory})   
\label{intro4}
\end{equation}
in {\em complete agreement} with the value 
\begin{math} \alpha_{\rm exp}\approx 2.7 \end{math} measured 
in cosmic ray experiments. The theory is based on computing 
the entropy per particle emitted from a cosmic ray source to 
be described more fully in what follows.  

In Sec.\ref{efp} we consider the evaporation of nucleons 
from a low temperature Landau-Fermi liquid droplet excited by 
an external collision. If one knows the entropy per nucleon 
\begin{math} \Delta s \end{math}  in the excited 
nuclear state, then the heat of evaporation 
\begin{math} q_{\rm vaporization}=T\Delta s \end{math} 
determines the energy distribution of vaporized nucleons employing  
the activation probability 
\begin{equation}
P(E)=\exp \big(-\Delta s(E)/k_B \big).  
\label{intro5}
\end{equation}
The quasi-particle entropy for a non-relativistic nucleon liquid 
droplet is reviewed below.

In Sec.\ref{ecrp}, we propose as the {\em sources} of cosmic 
rays, {\em the evaporating stellar winds from neutron star surfaces}. 
The quantum hadronic dynamical models of nuclear liquids have 
been a central theoretical feature of nuclear 
matter\cite{Walecka:1996,Serot:1986,Scalone:1999}. The theory of 
quantum hadronic matter is modeled in the main as an effective 
Bose (collective meson) theory. These models involve condensed 
scalar and vector mesons in about equal amounts. The scalar field 
may be thought to describe collective alpha nuclei embedded in the 
liquid while the vector field may be thought to describe deuteron nuclei 
embedded in the liquid. The neutron stars are evaporating through the 
surface from the nuclear matter within the bulk liquid into the ``vacuum'' 
or dilute gas. The gas contains the stellar wind of  cosmic rays emanating 
from the neutron star sources. In Sec.\ref{ce}, the exponent 
in Eqs.(\ref{intro2}) and (\ref{intro4}) are computed. In the 
concluding Sec.\ref{conc} the physics of the model is further discussed. 

\section{Evaporation of Fluid Particles \label{efp}}

The heat capacity per nucleon in a non-relativistic low 
temperature \begin{math} T \end{math} Landau-Fermi 
liquid drop is given by 
\begin{equation}
c=\frac{dE}{dT}=T\frac{ds}{dT}=\gamma T
\ \ \ {\rm as}\ \ \ T\to 0.
\label{epf1}
\end{equation} 
Eq.(\ref{epf1}) implies an excitation energy 
\begin{math} E=(\gamma /2)T^2 \end{math} and an 
entropy \begin{math} \Delta s=\gamma T \end{math} 
so that 
\begin{equation}
\Delta s(E)=\sqrt{2\gamma E}=k_B\sqrt{\frac{E}{E_0}}\ .
\label{epf2}
\end{equation} 

When a large nuclear matter nucleus is excited by an external 
collision or otherwise caused to be at a nonzero temperature
the energy distribution of nucleons that are evaporated 
is given by Eqs.(\ref{intro5}) and (\ref{epf2}) imply 
\begin{equation}
P(E)=\exp \left(-\sqrt{\frac{E}{E_0}}\right).
\label{epf3}
\end{equation} 
Let us now consider cosmic energy sources.

\section{Cosmic Ray Evaporation \label{ecrp}}

The structure of neutron stars consists of a big nuclear 
droplet\cite{Haensel:2007,Jin:2003} facing a very dilute gas, 
{\em i.e.} the ``vacuum''.  Neutron stars differ from being
simply very large nuclei in that most of their binding is 
gravitational rather than nuclear, but, the droplet model
of nuclear model should still offer a good description of
nuclear matter near the surface where it can evaporate.

The evaporation is from the effective fields in the form of scalar nuclei 
\begin{math} ^4He \equiv \alpha  \end{math} and vector deuterons 
\begin{math} ^2H \equiv d  \end{math}. Note that from neutron
stars we will not be in the low temperature regime discussed in 
the previous section, but rather at high temperatures where the
heat capacity goes to a constant.

What evaporates from the neutron star via scalar and vector fields 
are then energetic protons and alpha particles along with other 
nuclei to a much lesser extent. Deuterons are only weakly bound
and would be expected to photodissociate on background photons 
present throughout space rapidly to produce protons and
neutrons which in turn would produce protons when they decay.
In any event, experimentally, of course, at high energies discrimination of light
nuclei and protons is experimentally difficult.
The energy distribution used in the argument that follows comes directly
from the entropy of scalar (spin zero) and vector (spin 1) combinations of baryons. 

\section{The Power Law Exponent \label{ce}}

Ultra relativistic particles are virtually massless and thus have a density 
of energy states \begin{math} \propto \epsilon^2  \end{math}. The mean 
energy per particle then obeys 
\begin{equation}
E=\frac{\int_0^\infty f(\epsilon)\epsilon^3d\epsilon}
{\int_0^\infty f(\epsilon)\epsilon^2d\epsilon}
\label{ce1}
\end{equation}
wherein the Bose distribution is 
\begin{equation}
 f(\epsilon)=\frac{1}{e^{\epsilon/k_BT}-1}\ .
\label{ce2}
\end{equation}
Employing the Gamma  and zeta functions 
\begin{eqnarray}
 \Gamma (s)=\int_0^\infty e^{-x} x^s \left[\frac{dx}{x}\right],
 \nonumber \\ 
 \Gamma (s) \zeta (s)=\int_0^\infty 
 \left[\frac{x^s}{e^x - 1}\right] 
 \left[\frac{dx}{x}\right],
\label{ce3}
\end{eqnarray}
yields the mean energy 
\begin{equation}
E=cT \ \ \ {\rm wherein} 
\ \ \ c=k_B\left[\frac{\Gamma(4)\zeta (4)}
{\Gamma(3)\zeta(3)}\right].  
\label{ce4}
\end{equation}
The heat capacity per evaporated boson is thereby 
\begin{equation}
c= \alpha k_B =T\frac{ds}{dT}=\frac{dE}{dT}   
\label{ce5}
\end{equation}
wherein \begin{math} \alpha \end{math} is given in 
Eq.(\ref{intro2}). From Eqs.(\ref{ce4}) and (\ref{ce5}), 
the entropy obeys 
\begin{equation}
s(E)=\alpha k_B\ln \left(\frac{E}{E_0}\right).
\label{ce7}
\end{equation}
The power law energy distribution follows from 
Eqs.(\ref{ce7}) and (\ref{intro5}), 
\begin{equation}
P(E) \propto  \left(\frac{E_0}{E}\right)^\alpha 
\label{ce8}
\end{equation}
which is the central result of this work.

\section{Conclusions \label{conc}}

We have explored the hypothesis that cosmic rays are emitted from 
the surfaces of neutron stars. The cosmic rays themselves are in 
the stellar atmospheric winds blowing away from the neutron star 
source. The cosmic rays are equivalently nuclei which are evaporating 
from the bulk of the neutron star. On this basis, we find 
an inverse power in the energy distribution with a power law exponent of 
\begin{math} \alpha = 2.701178 \end{math} in excellent agreement 
with experimental data. The method of computing the energy distribution 
of the evaporated cosmic rays is closely analogous to those employed 
by Landau and Fermi for evaporation of nucleons from the Bohr-Mottelson 
liquid drop model. The correct exponent predicted by the model is 
satisfactory and encouraging. Further work is in progress with more
refined models.

\section{Acknowledgements}

J.S.  thanks the National Science Foundation for its support via NSF grant
PHY-1205845.

\vfill \eject

\end{document}